\documentclass[10pt]{article}
\usepackage{graphicx}
\usepackage{amsmath}

\def\Title#1{\begin{center} {\Large #1 } \end{center}}
\def\Author#1{\begin{center}{ \sc #1} \end{center}}
\def\Address#1{\begin{center}{ \it #1} \end{center}}

\newcommand\pubblock{\rightline{\begin{tabular}{l} Proceedings of the Fifth Annual LHCP\\ \pubnumber\\
         \pubdate  \end{tabular}}}

\newenvironment{Abstract}{\begin{quotation} \begin{center} 
             \large ABSTRACT \end{center}\bigskip 
      \begin{center}\begin{large}}{\end{large}\end{center} \end{quotation}}

\newenvironment{Presented}{\begin{quotation} \begin{center} 
             PRESENTED AT\end{center}\bigskip 
      \begin{center}\begin{large}}{\end{large}\end{center} \end{quotation}}

\def\beq{\begin{equation}}
\def\eeq#1{\label{#1}\end{equation}}
\def\eeqn{\end{equation}}

\def\beqa{\begin{eqnarray}}
\def\eeqa#1{\label{#1}\end{eqnarray}}
\def\eeqan{\end{eqnarray}}

\let\bar=\overbar

\def\Dslash{\not{\hbox{\kern-4pt $D$}}}
\def\dslash{\not{\hbox{\kern-2pt $\del$}}}

\def\msb{{\bar{\ssstyle M \kern -1pt S}}}

\usepackage{color}

\newcommand\pubnumber{ ATL-PHYS-PROC-2017-XXX }

\newcommand\pubdate{\today}

\def\affiliation{
On behalf of the ATLAS, LHCb and CMS Collaborations, \\
Department of Physics \\
University of Wisconsin-Madison, Madison, WI 53706-1390, USA}

\begin{document}

\large
\begin{titlepage}
\pubblock

\vfill
\Title{  ELECTROWEAK MEASUREMENTS AT THE LHC  }
\vfill

\Author{ ALEXANDER SAVIN  }
\Address{\affiliation}
\vfill
\begin{Abstract}

A set of selected electroweak measurements from the LHC experiments is discussed. Results on forward-backward asymmetry in production of the Drell-Yan events in both dielectron 
and dimuon decay channels are presented together with results on the effective mixing angle measurements.
 Angular coefficients measured in the Z boson production are compared with theoretical predictions. Electroweak production of the vector bosons in association with two jets is presented. 
\end{Abstract}
\vfill

\begin{Presented}
The Fifth Annual Conference\\
 on Large Hadron Collider Physics \\
Shanghai Jiao Tong University, Shanghai, China\\ 
May 15-20, 2017
\end{Presented}
\vfill
\end{titlepage}
\def\thefootnote{\fnsymbol{footnote}}
\setcounter{footnote}{0}
%

\normalsize 


\section{Introduction}

The electroweak (EWK) measurements are playing an important role at the LHC. The measured cross sections
allow better understanding of the standard model (SM) predictions and   
 backgrounds to the searches beyond the SM. Recently available predictions at the 
next-to-next-to-leading order (NNLO) in QCD and next-to-leading order (NLO) in EWK require
for comparisons high-precision measurements with well understood sources of the 
experimental and theoretical systematic uncertainties.
The global SM parameters: vector boson masses, $sin \theta_W$, $\alpha_{em}$ can also be
 measured thus providing important input to the models and global fits.

These proceedings cover only few selected EWK measurements performed by ATLAS~\cite{Aad:2008zzm}, 
LHCb~\cite{Alves:2008zz,Aaij:2014jba}, and CMS~\cite{Chatrchyan:2008aa} collaborations. The measurements are related mainly to single vector boson production,  more topics are discussed in other contributions that 
 were presented at the same conference.

\section{Forward-backward asymmetry in Drell-Yan production}

\begin{figure}[htb]
\centering
\includegraphics[height=2in]{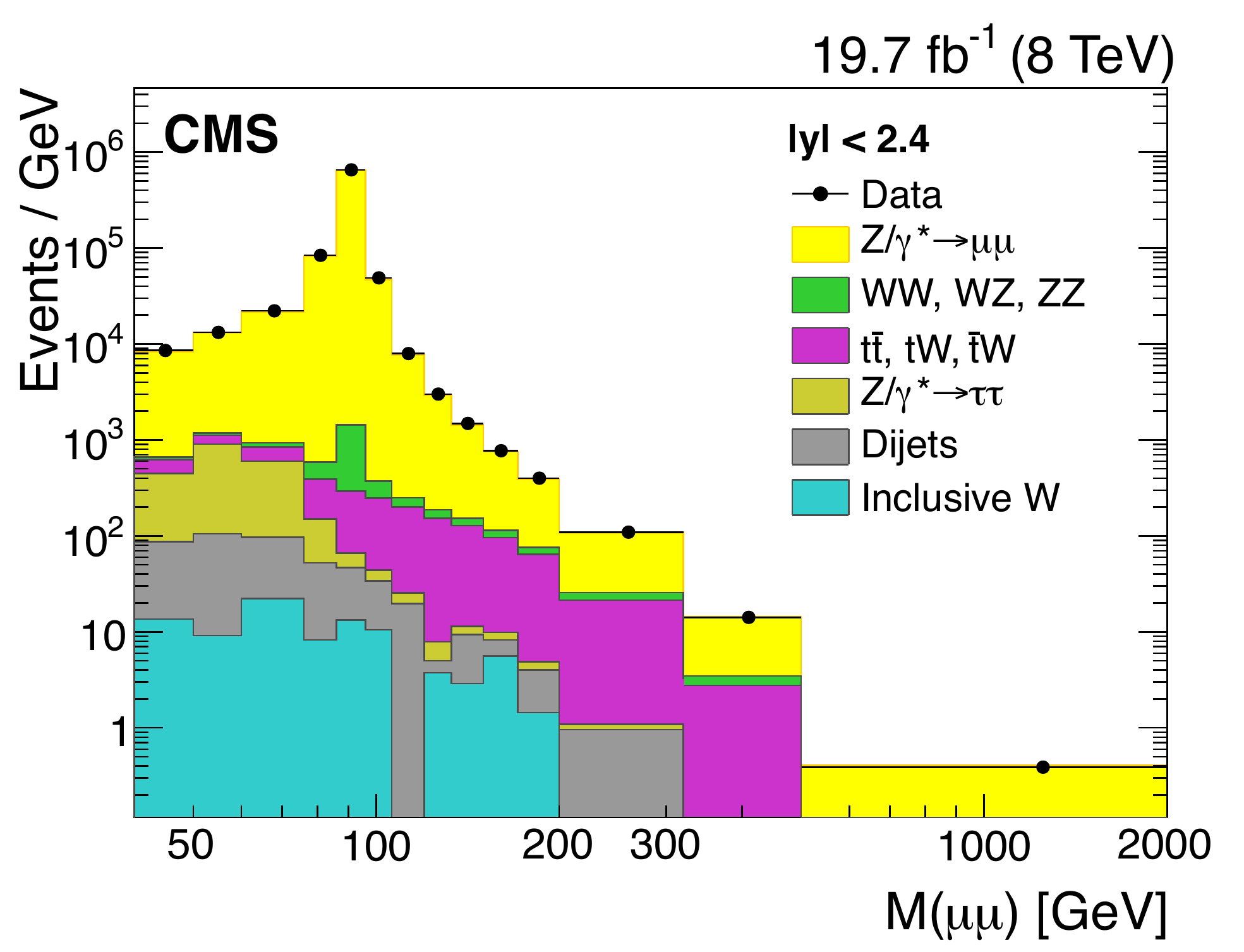}
\includegraphics[height=2in]{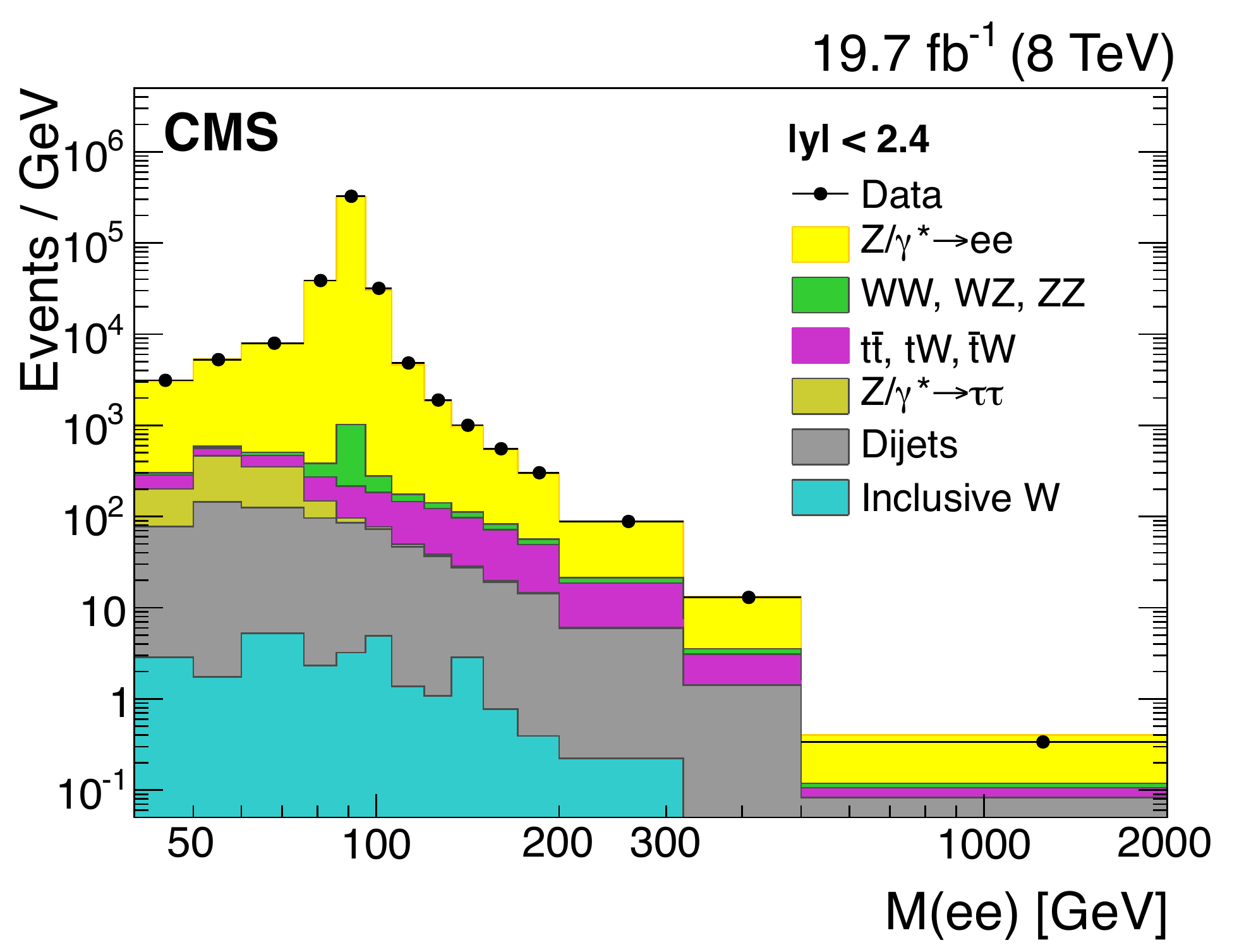}
\caption{The dilepton reconstructed mass distributions for muon (left) 
and electron (right)  
decay channels, for events with $|y| < 2.4$ ~\cite{Khachatryan:2016yte}. }
\label{fig:figure1}
\end{figure}

\begin{figure}[htb]
\centering
\includegraphics[height=2in]{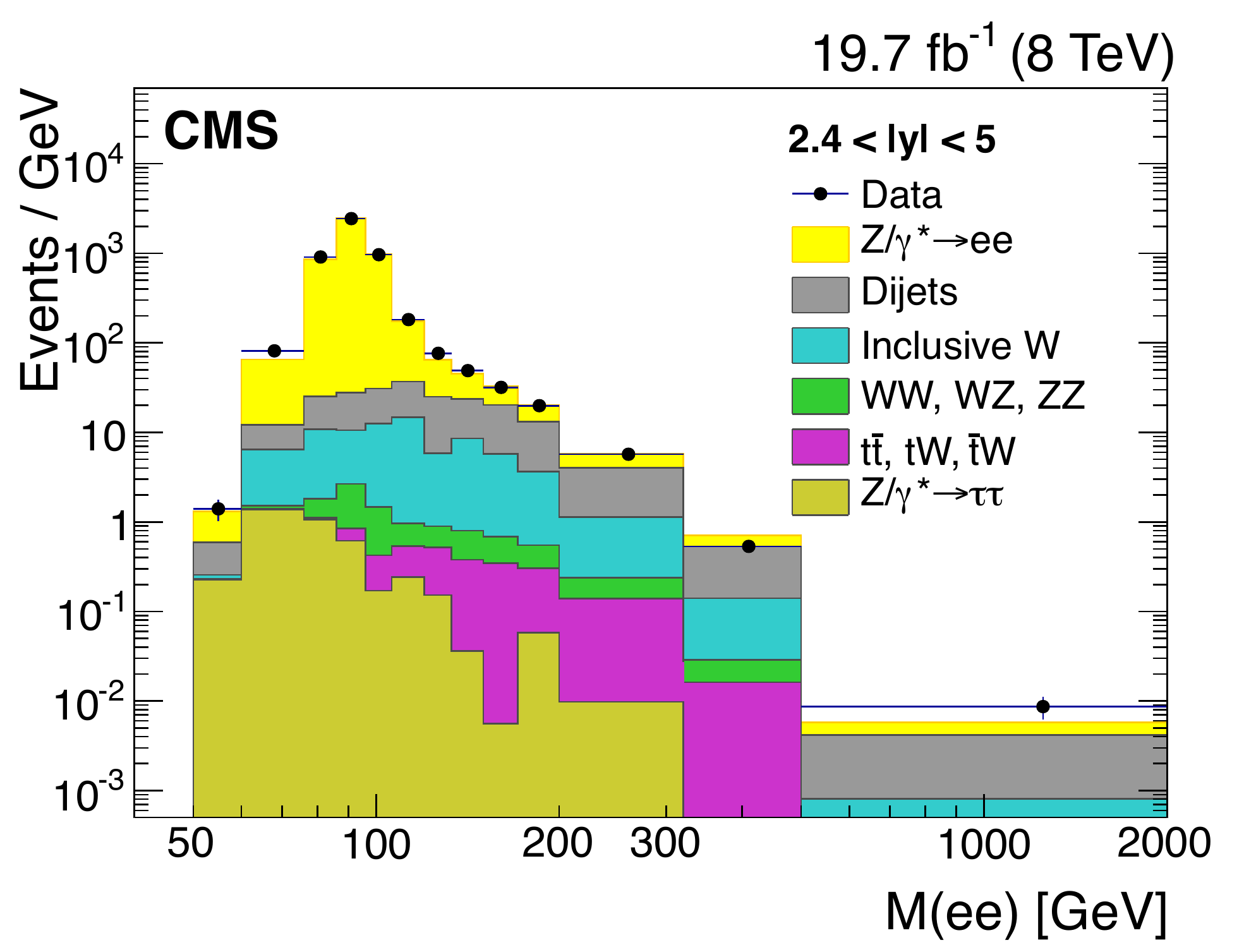}
\includegraphics[height=2in]{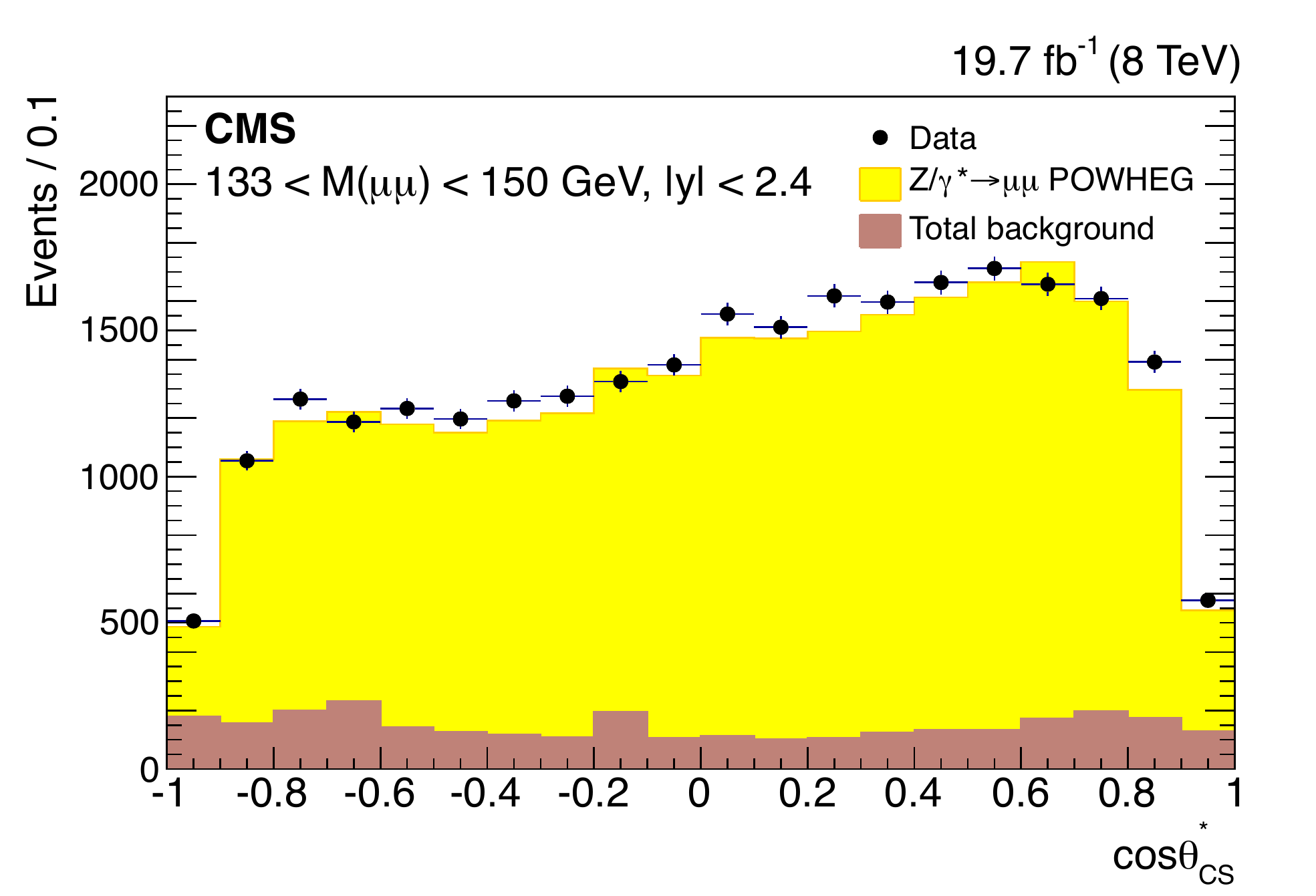}
\caption{ (left) The dilepton mass distributions for 
electron decay channels, for events with $2.4 < |y| < 5$. 
(right) A representative
 $\cos {\theta^*_{CS}}$ distribution for dimuon channel for $\mu\mu$ mass range
133 -- 150 GeV ~\cite{Khachatryan:2016yte}.}
\label{fig:figure2}
\end{figure}

The presence of both vector and axial-vector couplings of electroweak bosons to fermions lead to
a forward-backward asymmetry $A_{FB}$ in the production of Drell-–Yan lepton pairs.
The $A_{FB}$ is defined as $A_{FB} = \frac {\sigma_F - \sigma_B} {\sigma_F + \sigma_B}$, where 
$\sigma_{F(B)}$ is the total cross section for the forward ($cos \theta^* > 0$) and backward ($cos \theta^* < 0$) events.
To reduce the uncertainties due to the transverse momentum of the incoming quarks, the measurements use the Collins-Soper (CS)
frame~\cite{CollinsSoper}. 
In the recent CMS measurement~\cite{Khachatryan:2016yte} at $\sqrt{s} = 8$ TeV the DY events were detected in decays to electron
and muon pairs, with lepton transverse momentum, $p_T$, above 20 GeV and pseudorapidity $|\eta| < 2.4$ for muons, 
while for electrons the region was extended up to $|\eta| < 5$ by using forward hadron 
(HF) calorimeter for the electrons identification.  
The measurement is performed as a function of dilepton mass in bins of rapidity , $y$, 
of the dilepton system. Figure~\ref{fig:figure1} shows the dilepton mass distributions for muon and 
electron decay channels, for events with $|y| < 2.4$. The extended region in $y$ for electron pairs is 
shown in Fig.~\ref{fig:figure2} (left). The data are well described by the simulation convoluted with 
data-driven background estimates. The backgrounds
are relatively small. The major experimental uncertainties arise from the electron and muon energy corrections 
and from the unfolding procedure. The mass resolution in the forward region is not as good as in the central one, but this region 
is important since the ambiguity of the quark direction is lower at higher $y$ and the dilution 
of $A_{FB}$ is therefore smaller.

The $A_{FB}$ measurement is performed as a function of dilepton mass in bins of rapidity. The shape of  $\cos {\theta^*_{CS}}$ changes with the mass, 
one representative distribution for the dimuon channel for $\mu\mu$ mass range 133 -- 150 GeV is shown in 
Fig.~\ref{fig:figure2} (right). The stacked histograms represent the sum of the background contribution
and signal. The data are well described by the expected distributions.

\begin{figure}[htb]
\centering
\includegraphics[height=2in]{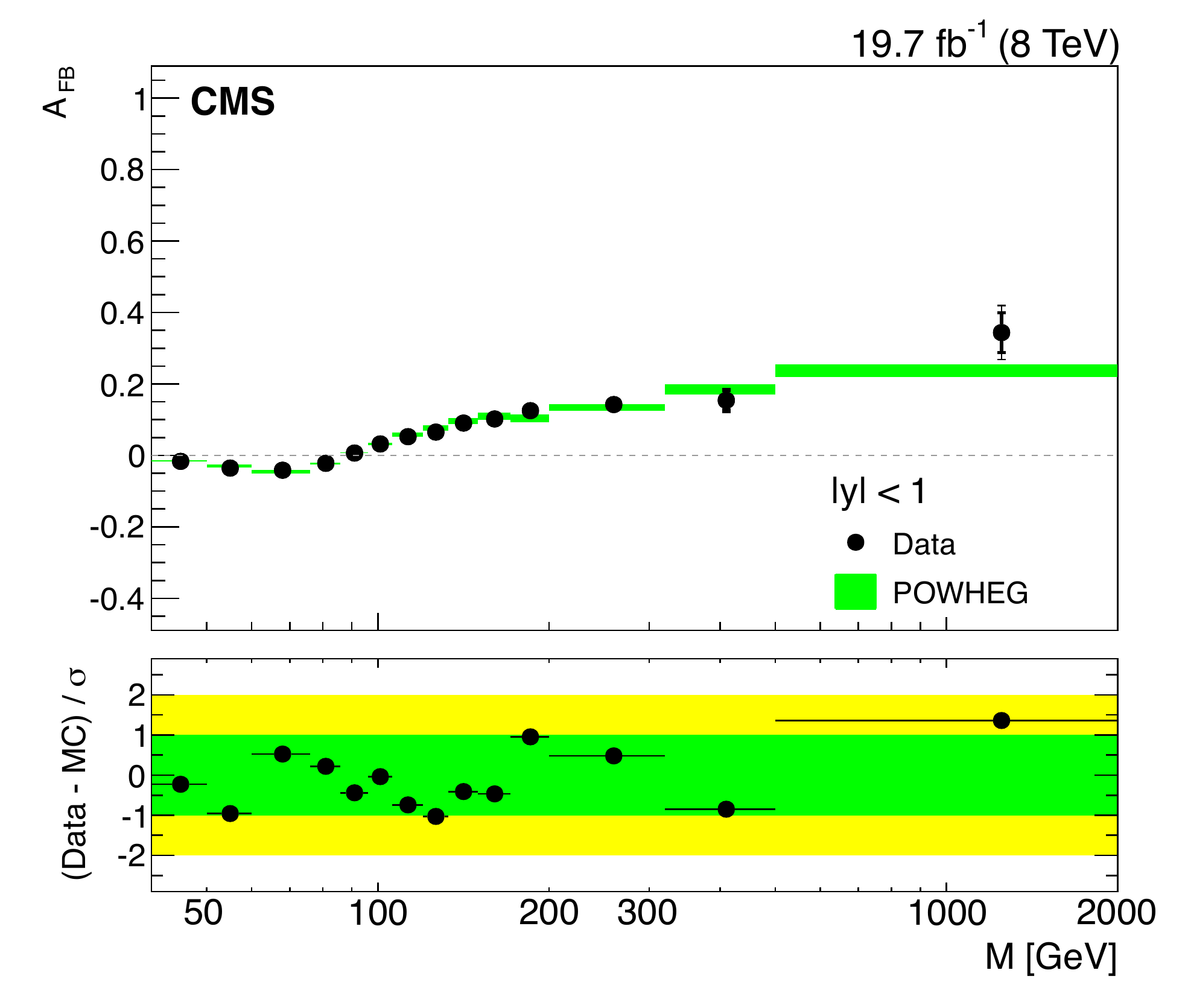}
\includegraphics[height=2in]{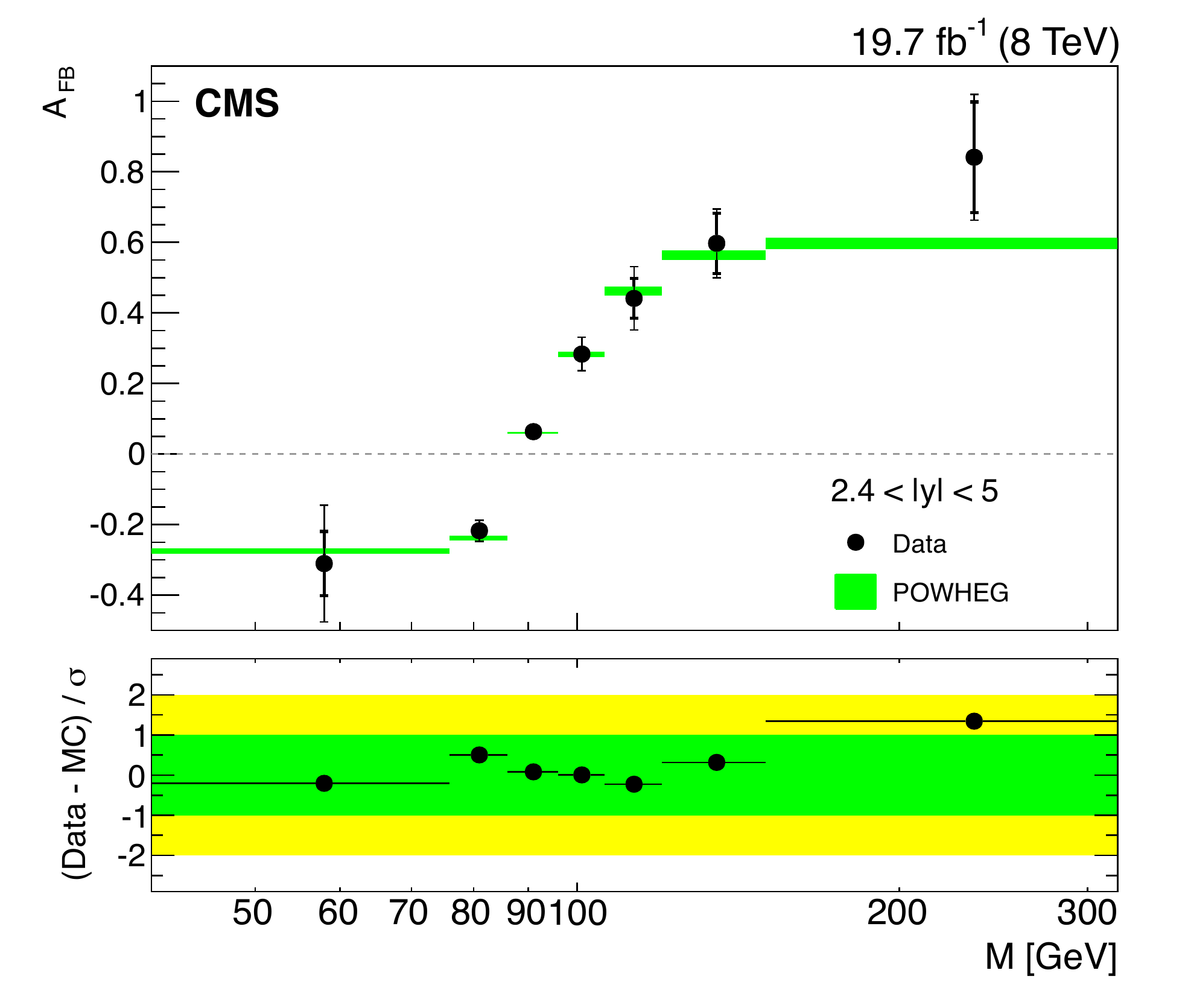}
\caption{The combined dielection and dimuon unfolded $A_{FB}$ distributions
in the central rapidity region $|y| <1$ (left) and in the forward region for
dielectron decay channel (right) ~\cite{Khachatryan:2016yte}. }
\label{fig:figure3}
\end{figure}

The combined dielection and dimuon unfolded $A_{FB}$ distributions
in the central rapidity region $|y| <1$ and in the forward region for
dielectron decay channel are shown in Fig.~\ref{fig:figure3}. The measured
distributions agree well with the POWHEG predictions. Because $A_{FB}$ in the
forward rapidity region is less diluted, the measured $A_{FB}$ quantity is closer to the 
parton-level asymmetry after the unfolding process, than it is in the central rapidity region.

\section{Effective mixing angle measurements}

Measurement of the backward-forward asymmetry can be used for extraction of the
effective mixing angle. Such measurement was performed by all three experiments
ATLAS~\cite{Aad:2015uau}, LHCb~\cite{Aaij:2015lka} and CMS~\cite{Chatrchyan:2011ya}, only recent LHCb result is discussed here in detail. 

\begin{figure}[htb]
\centering
\includegraphics[height=2in]{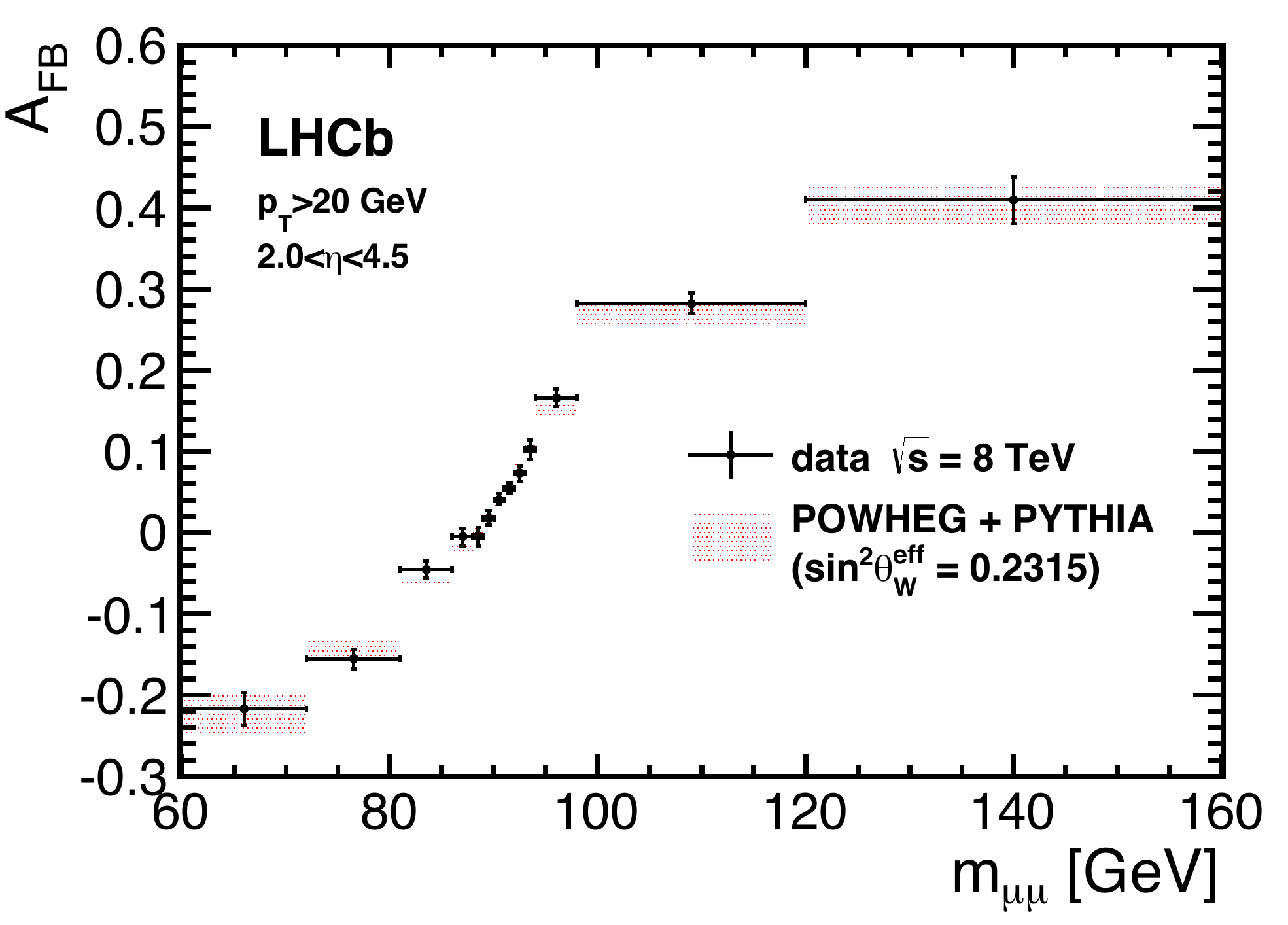}
\includegraphics[height=2in]{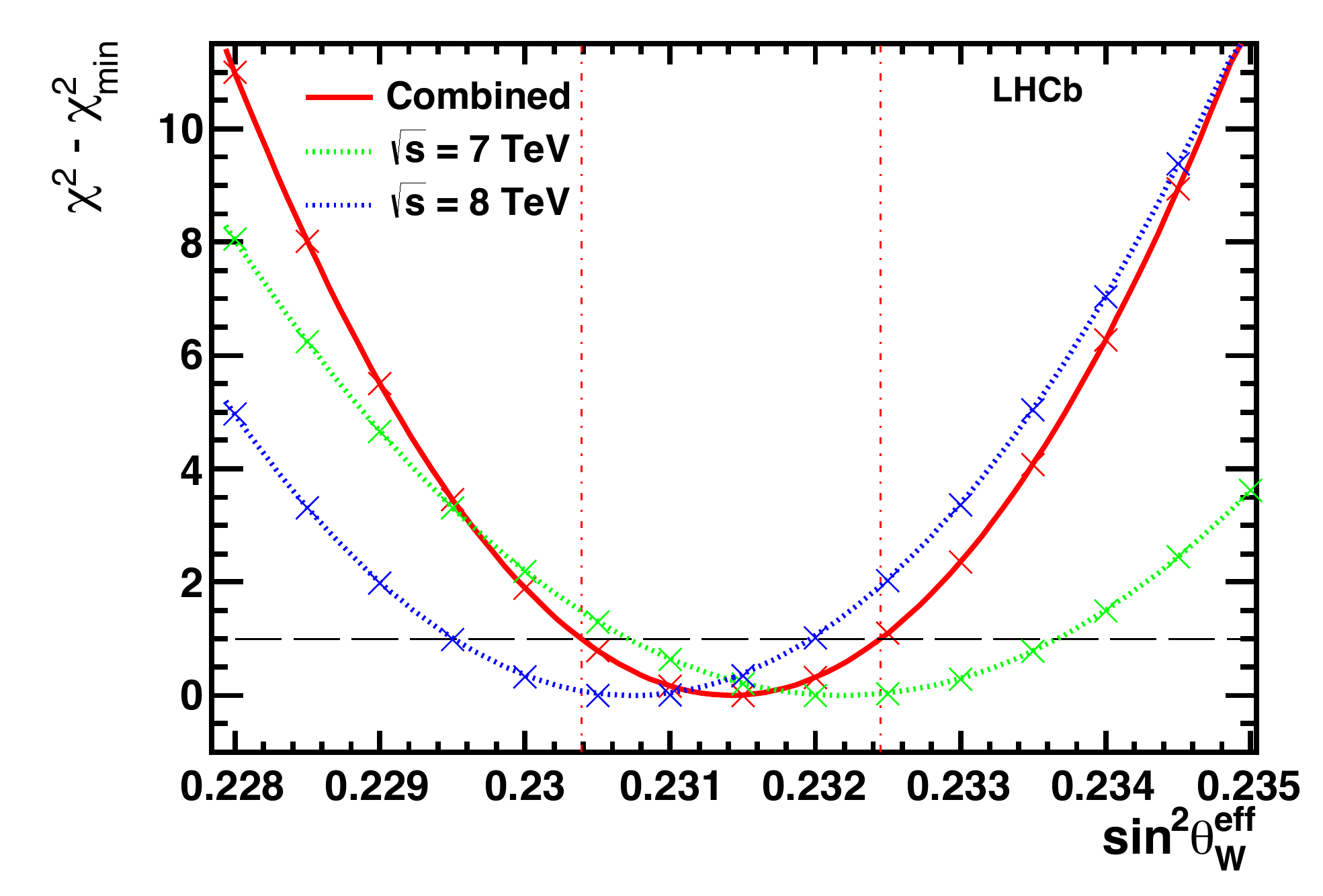}
\caption{(Left) The $A_{FB}$ as a function of the dimuon invariant mass. (Right) Difference between the $\chi^2$ and the minimum $\chi^2$ obtained by comparing the measured and predicted $A_{FB}$ distribution for different values of $\sin^2{\theta^{eff}_W}$~\cite{Aaij:2015lka}.  
 }
\label{fig:figure4}
\end{figure}

The LHCb potentially has higher power for measuring the effective mixing angle, than ATLAS and CMS, since it naturally collects events in the forward region,
$2 < \eta < 5$.
Figure~\ref{fig:figure4} left shows the $A_{FB}$ as a function of dimuon mass
as measured in LHCb. Both muons are required to be within $2.0 < \eta < 4.5$ and have transverse momentum greater than 20 GeV. The measurements are performed
with two data samples, at $\sqrt{s} = 7$ and 8 TeV, with luminosities of 1 and 2 fb$^{-1}$ respectively.   
The $A_{FB}$ as a function of the dimuon invariant mass is compared with several sets of SM predictions generated with POWHEG for values of $\sin^2{\theta^{eff}_W}$ ranging from 0.22 to 0.24. The Z-boson mass and electromagnetic 
coupling constant were fixed to their PDG values, NNPDF2.3 PDF set~\cite{NNPDF}  was used with the strong coupling constant of 0.118. 
The agreement between data and predictions is quantified using $\chi^2$ value, taking into account statistical, systematic and theoretical uncertainties, and correlations between mass bins. A quadratic function is fitted
to the $\chi^2$ as shown in Fig.~\ref{fig:figure4} right.
The interval in $\sin^2{\theta^{eff}_W}$ corresponding to variation of one unit in $\chi^2$ is quoted as the uncertainty. 
Combination of 7 and 8 TeV results obtained by calculating the full covariance matrix for all uncertainties yields the value $\sin^2{\theta^{eff}_W} = 0.23142 \pm 0.00073 (stat) \pm 0.00052 (sys) \pm 0.00056 (theo)$. 

\begin{figure}[htb]
\centering
\includegraphics[height=2.1in]{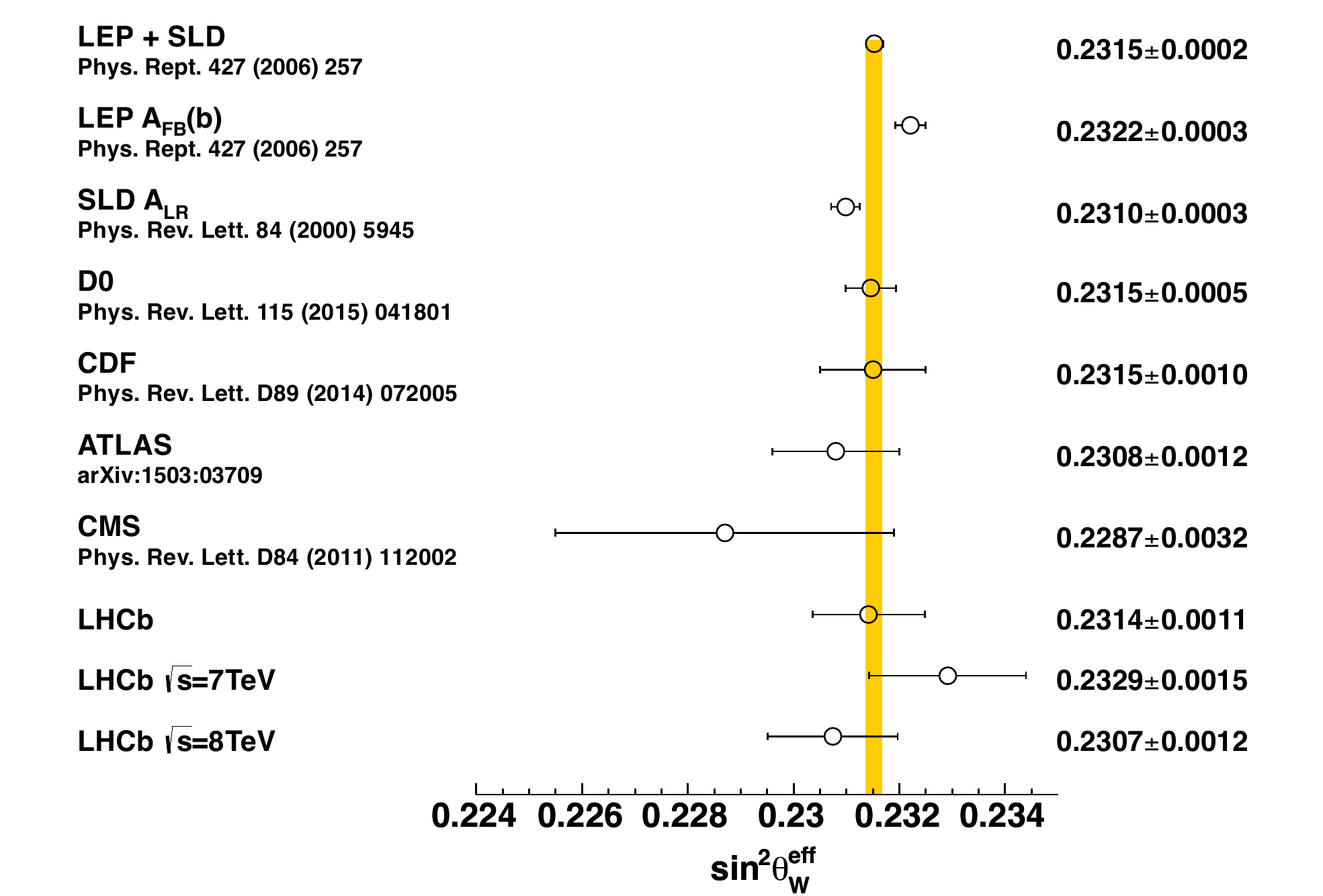}
\caption{A comparison of the $\sin^2{\theta^{eff}_W}$
 measurement at LHCb and other experiments. The combined LEP and SLD measurement is indicated by the vertical yellow band~\cite{Aaij:2015lka}. 
 }
\label{fig:figure5}
\end{figure}

A comparison between the $\sin^2{\theta^{eff}_W}$ results obtained by different experiments
is shown in Fig.~\ref{fig:figure5}. The LHCb result agrees well with the world average and LHC results.

\section{Angular coefficients in Z-bosons events}

The general structure of the lepton angular distribution in the boson rest frame is given by

\begin{equation}
\label{eq:x-section}
\begin{split}
\frac {d^2 \sigma } {d\cos\theta^{*}d\phi^{*}}
\propto \Bigl[(1+\cos^2\theta^{*}) +A_0 \frac{1}{2}(1-3\cos^2\theta^{*}) + A_1\sin(2\theta^{*})\cos\phi^{*} + A_2\frac{1}{2}\sin^2\theta^{*}\cos(2\phi^{*})
\\
 +A_3\sin\theta^{*}\cos\phi^{*} + A_4\cos\theta^{*} + A_5 \sin^2 \theta^* \sin(2\phi^*)+
A_6\sin(2\theta^*)\sin{\phi^*} + A_7\sin{\theta^*}\sin{\phi^*}\Bigr].
\end{split}
\end{equation}

where the $\theta^*$ is defined as in $A_{FB}$ measurement, and $\phi^*$ is the azimuthal angle.

\begin{figure}[htb]
\centering
\includegraphics[height=2.1in]{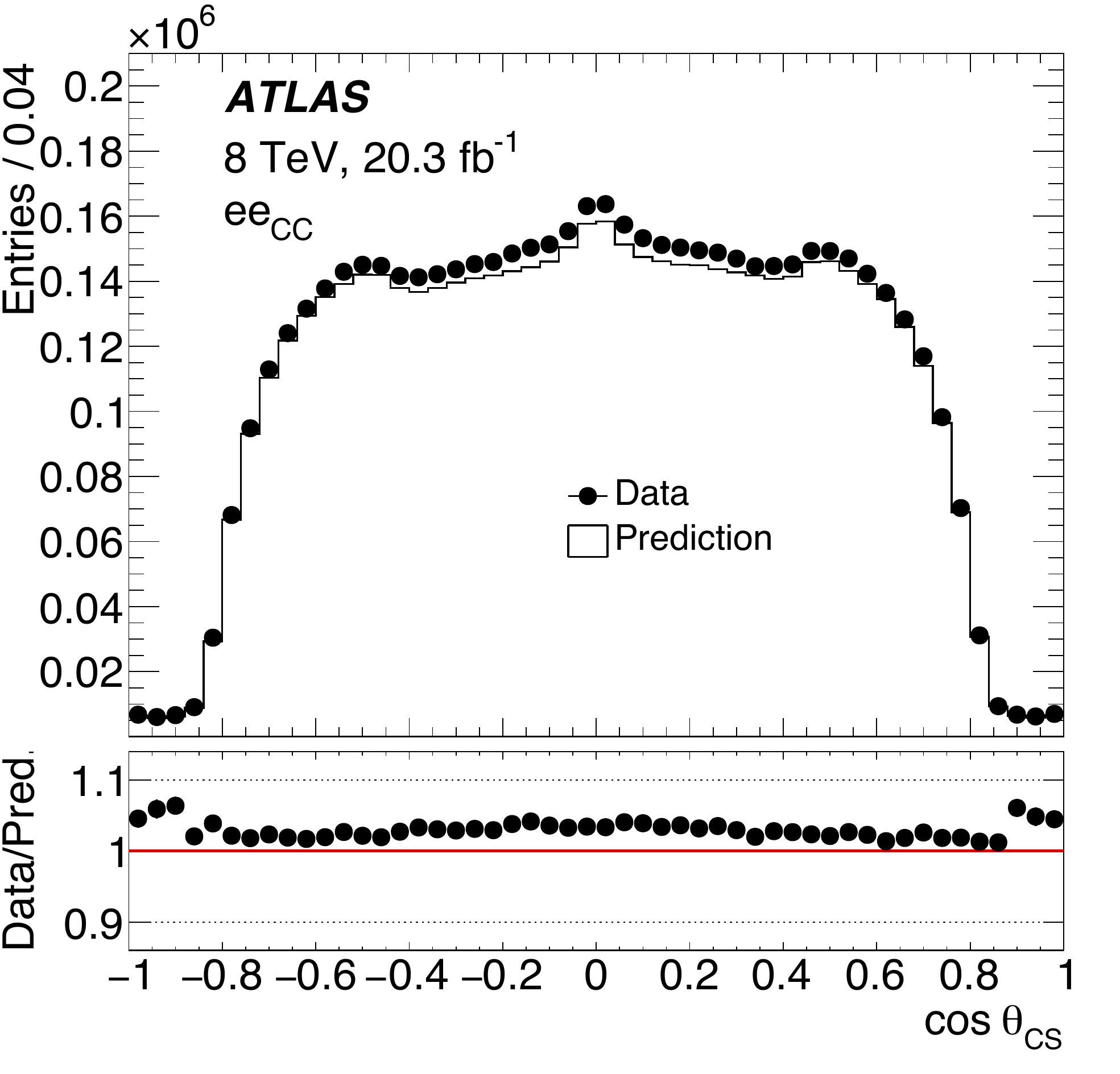}
\includegraphics[height=2.1in]{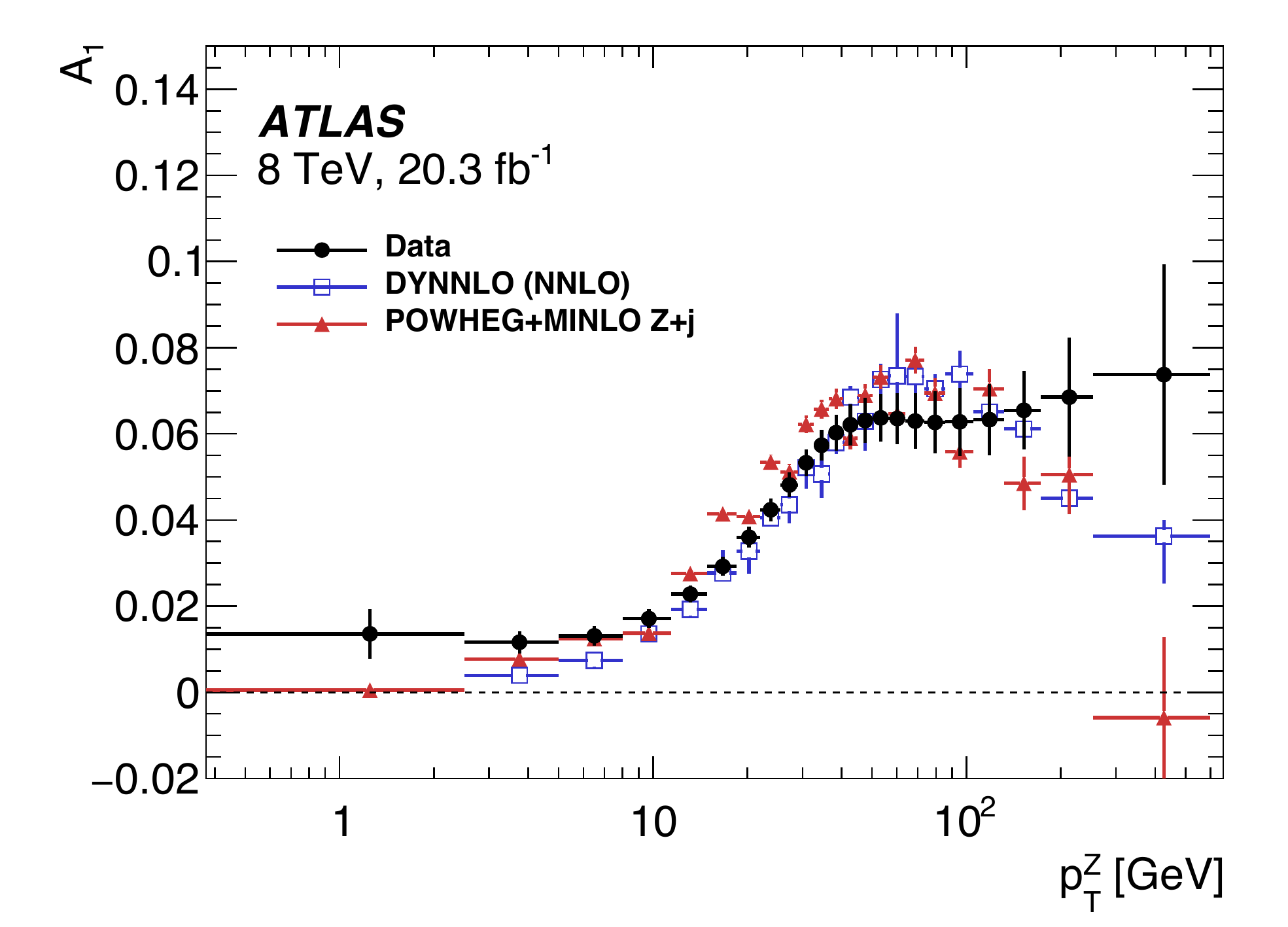}
\caption{(Left) The $\cos \theta_{cs}$ angular distributions for dielectron channel. In the panel the ratios of the data to the summed signal+background predictions is shown, the uncertainty bars on the points only include those from data statistics. (Right) Distribution of the angular coefficient $A_1$ as a function of $p^Z_T$. 
The results are compared to the DYNNLO and POWHEG MINLO predictions~\cite{Aad:2016izn}.
}
\label{fig:figure6}
\end{figure}

Parameters $A_{0}$, $A_{1}$, and $A_{2}$ are related to the polarization of the Z boson, whilst $A_{3}$ and $A_{4}$ are also sensitive to the V-A structure of the couplings of the leptons.
 All angular coefficients vanish as the Z boson transverse momentum  approaches zero
except for $A_{4}$, which is the electroweak parity violation term.
The Lam--Tung relation~\cite{Lam:1978zr} $A_0=A_2$ reflects 
the full transverse polarization of vector boson coupling to quarks, as well as rotational invariance.
Processes containing non-planar configurations (e.g., from higher order multi-gluon emission) smear the transverse polarization, leading to $A_{2} < A_{0}$.
The $A_5$, $A_6$ and $A_7$ coefficients appear first at NNLO in QCD.

In this section we discuss in detail only most recent ATLAS measurement~\cite{Aad:2016izn} that was performed at 8 TeV for Z-bosons decaying to both 
electron and muon pairs with transverse momenta of the leptons above 25 GeV and for the Z-boson mass in 
the range 80--100 GeV.

The measurement of the angular coefficients is performed in multiple fine bins of $p^Z_T$ 
and for a fixed dilepton mass window. Thus the data-simulation agreement in shape for these variables is less important,
but it is important  to 
verify qualitatively the level of agreement between data and MC simulation for the 
angular distributions, as shown in Fig.~\ref{fig:figure6} left. The data and MC distributions 
are not normalized to each other, resulting in normalization differences at the level of
 a few percent. The measurement of the angular coefficients is, however, independent of this difference.
The coefficients are extracted from the data by fitting templates to the reconstructed angular distributions. 
Each template is normalized by free parameters for its corresponding coefficient $A_i$, 
as well as an additional common parameter representing the unpolarised cross section. 
A likelihood is built from the nominal templates and the varied templates reflecting the systematic 
uncertainties. The muon and electron channels are combined through a likelihood multiplication.

\begin{figure}[htb]
\centering
\includegraphics[height=2.1in]{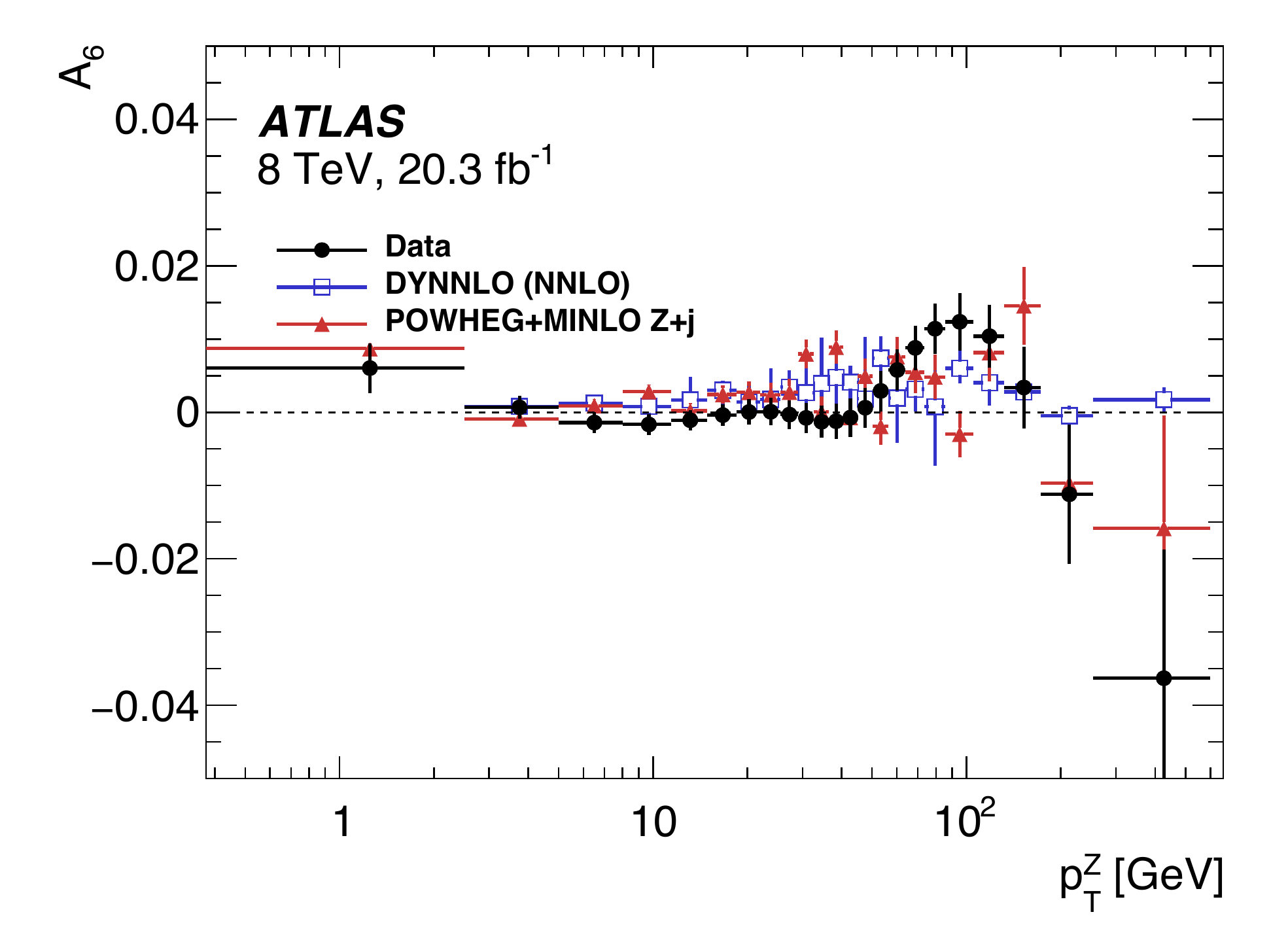}
\includegraphics[height=2.1in]{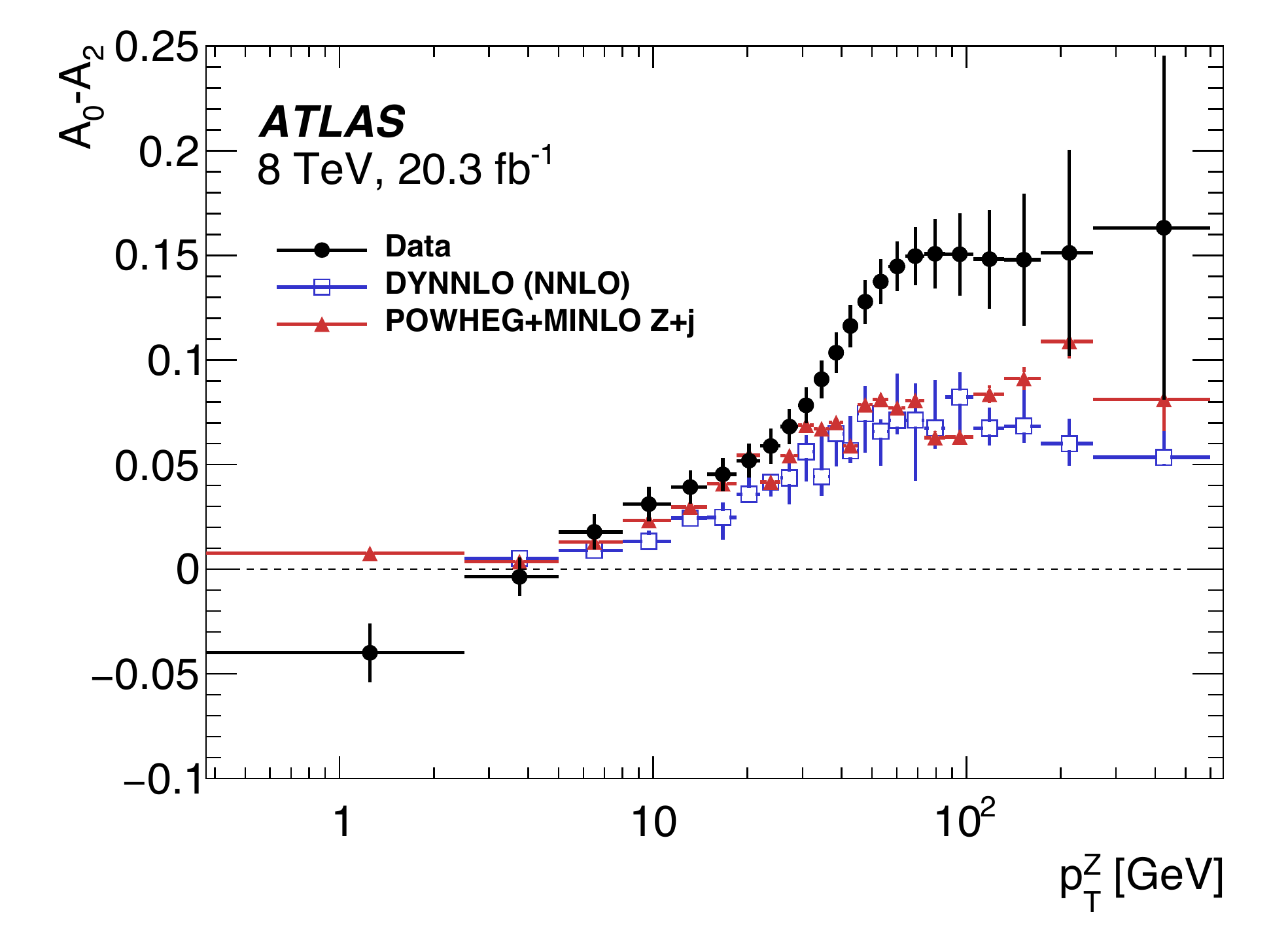}
\caption{Distributions of the angular coefficients $A_5$ (left) and $A_0-A_2$ (right) 
as a function of $p^Z_T$. The results from the measurements are compared to the DYNNLO and POWHEG MINLO predictions~\cite{Aad:2016izn}. 
 }
\label{fig:figure7}
\end{figure}

Figure~\ref{fig:figure6} (right) shows measured distribution of the angular coefficient $A_1$ as a function of $p^Z_T$
 compared to the DYNNLO and POWHEG MINLO predictions. The NNLO predictions describe the data with 
some small discrepancy towards high $p^Z_T$ values, still within uncertainties. Similar behavior
was observed also for other coefficients. For the first time at the LHC the $A_5$ -- $A_7$ coefficients
are measured as shown in Fig.~\ref{fig:figure7} left. As expected the Lam--Tung relation does not hold, 
but as one can see in Fig.~\ref{fig:figure7} right the NNLO predicts much smaller deviation from 
0 than observed in data.
ATLAS results agree well with the previous CMS measurement~\cite{Khachatryan:2015paa}, while extending it to measurement of 
the $A_5$ -- $A_7$ coefficients.

\section{Electroweak production of vector bosons in association with two jets}

Electroweak production of vector bosons is characterized by production of one, in case 
of W production, or two, in case of Z,  
leptons in the central part of the detector with two jets in backward/forward directions 
separated by a large rapidity gap. The major background to EWK process is W/Z+jj QCD production.
In the recent ATLAS Wjj analysis~\cite{Aaboud:2017fye} for $\sqrt{s}=7$ and 8 TeV, 
the electrons and muons are required to have $p_T > 25$ GeV,
two jets with $p_T > 80(60)$ GeV, separated by $\Delta |y| > 2$, in presence of missing
transverse energy (greater than 20 GeV) and transverse mass, $m_T > 40$ GeV. The signal
region should contain only one lepton in the central region and no jets.

\begin{figure}[htb]
\centering
\includegraphics[height=2.5in]{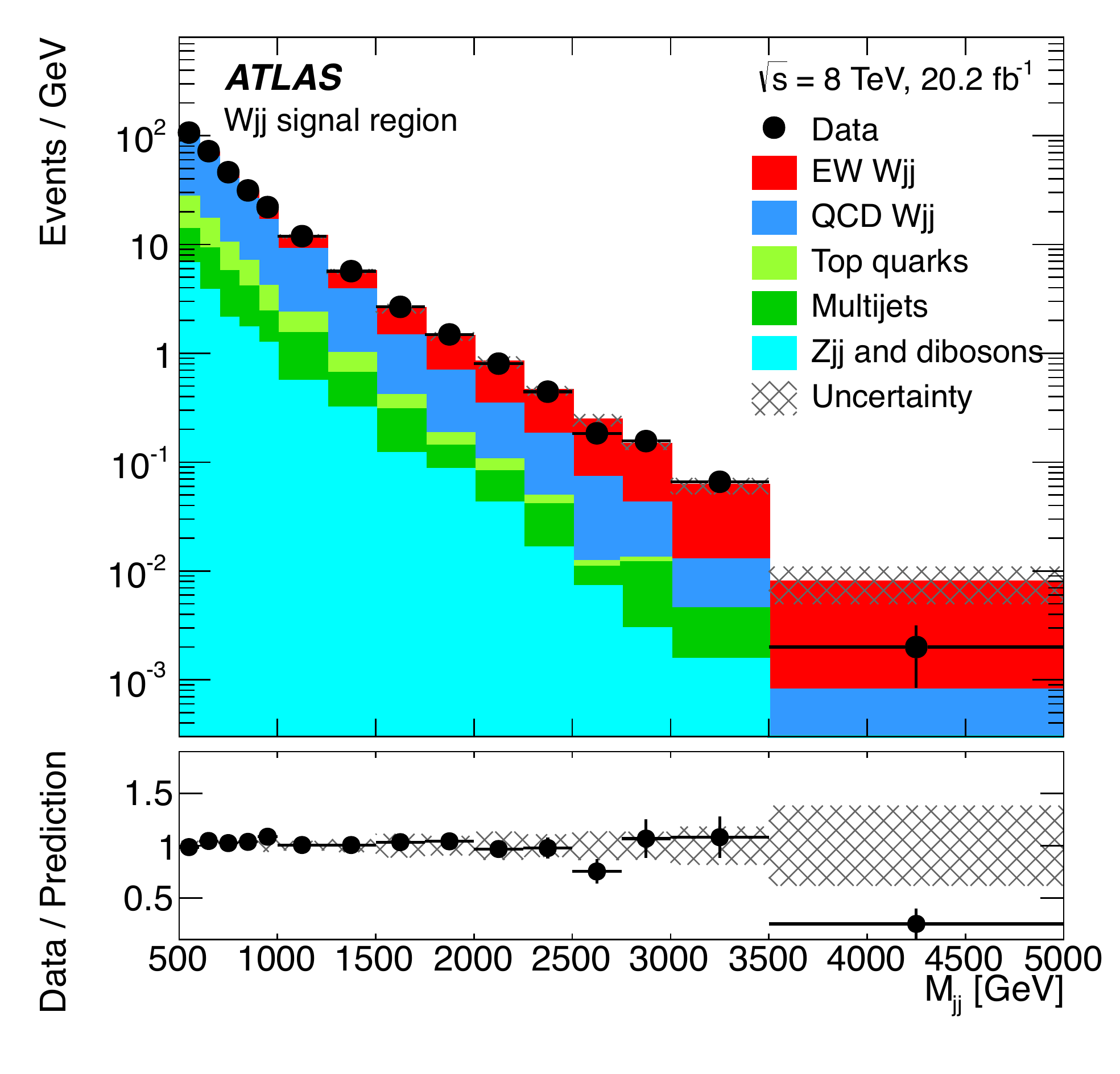}
\includegraphics[height=2.5in]{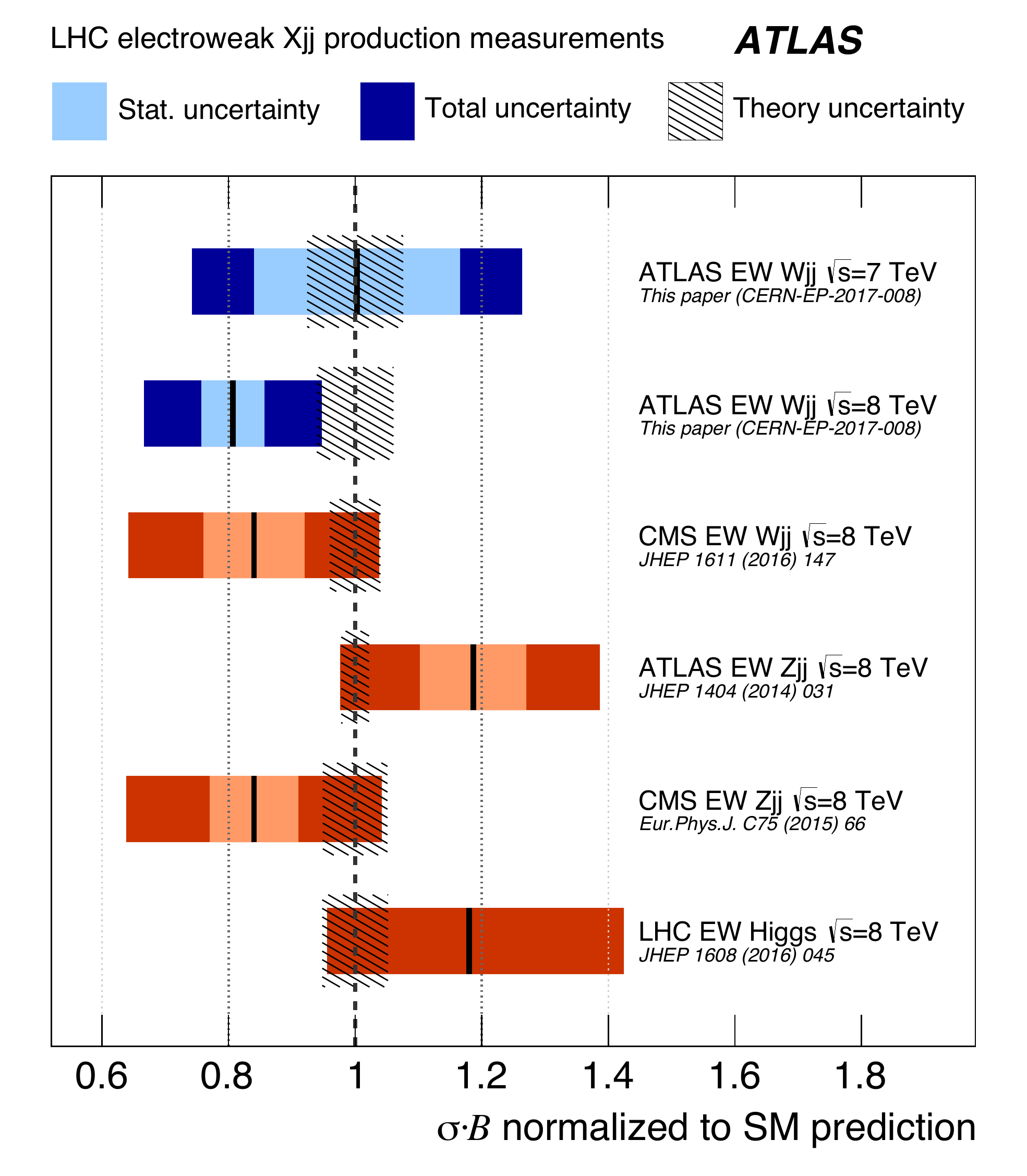}
\caption{(Left) 
Distribution of the dijet invariant mass for events in the signal region in 8 TeV data, 
after fitting for the yields of the individual Wjj processes. The bottom panel shows the ratio of 
data to predicted signal-plus-background yields. The shaded band centred at unity represents the 
statistical and experimental uncertainties summed in quadrature. 
(Right) Measurements of the cross section times branching fractions of electroweak production of a 
single W, Z, or Higgs boson at high dijet invariant mass, 
divided by the SM predictions 
(Powheg+Pythia8 for ATLAS, Madgraph+Pythia8 for CMS, and Powheg+Pythia8 for the LHC combination). 
The lighter shaded band (where shown) represents the statistical uncertainty of the measurement, 
the outer darker band represents the total measurement uncertainty. Theoretical uncertainties in the 
SM prediction are represented by the shaded region centred at unity~\cite{Aaboud:2017fye}. 
 }
\label{fig:figure8}
\end{figure}

\begin{figure}[htb]
\centering
\includegraphics[height=2.5in]{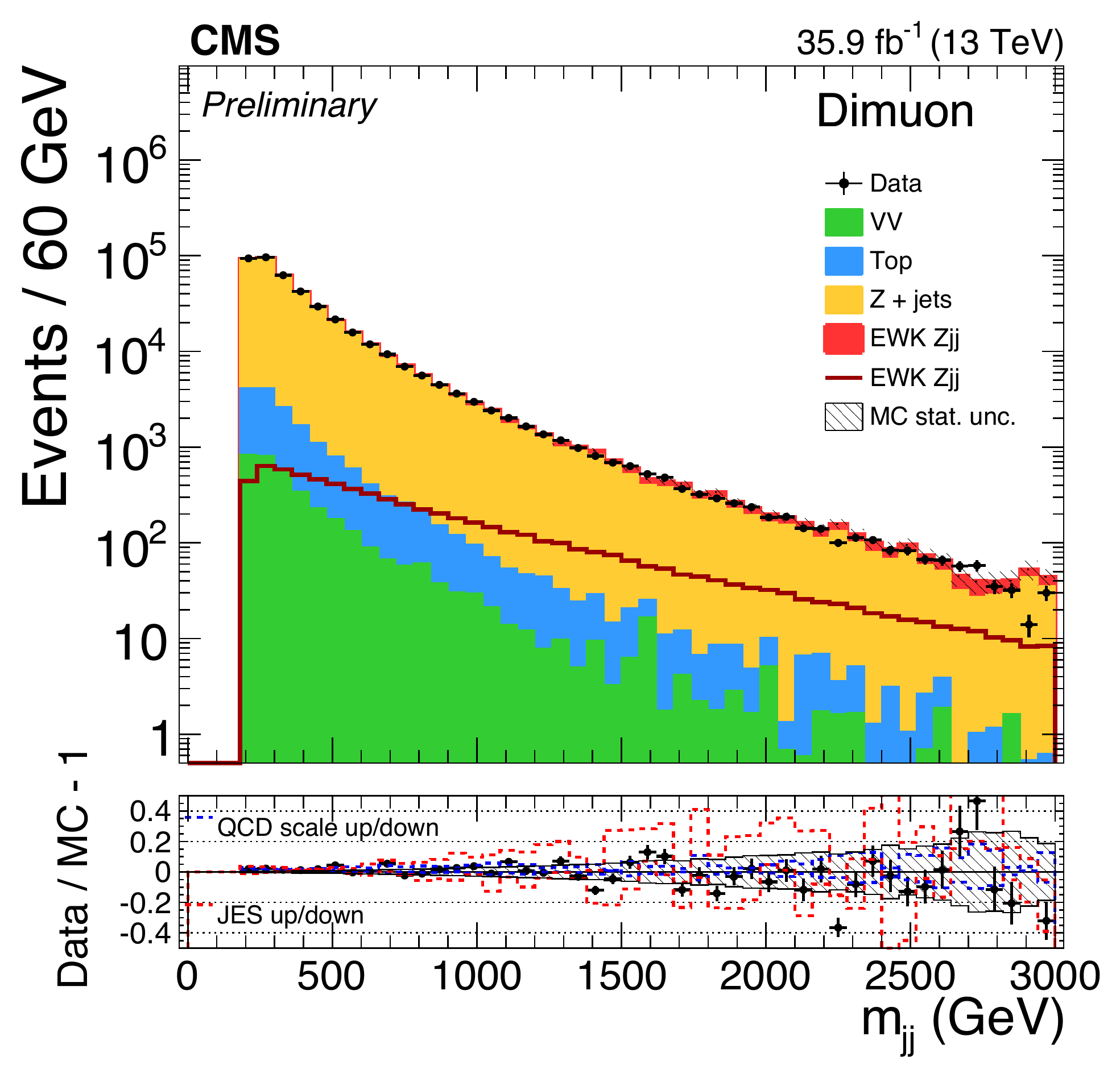}
\includegraphics[height=2.5in]{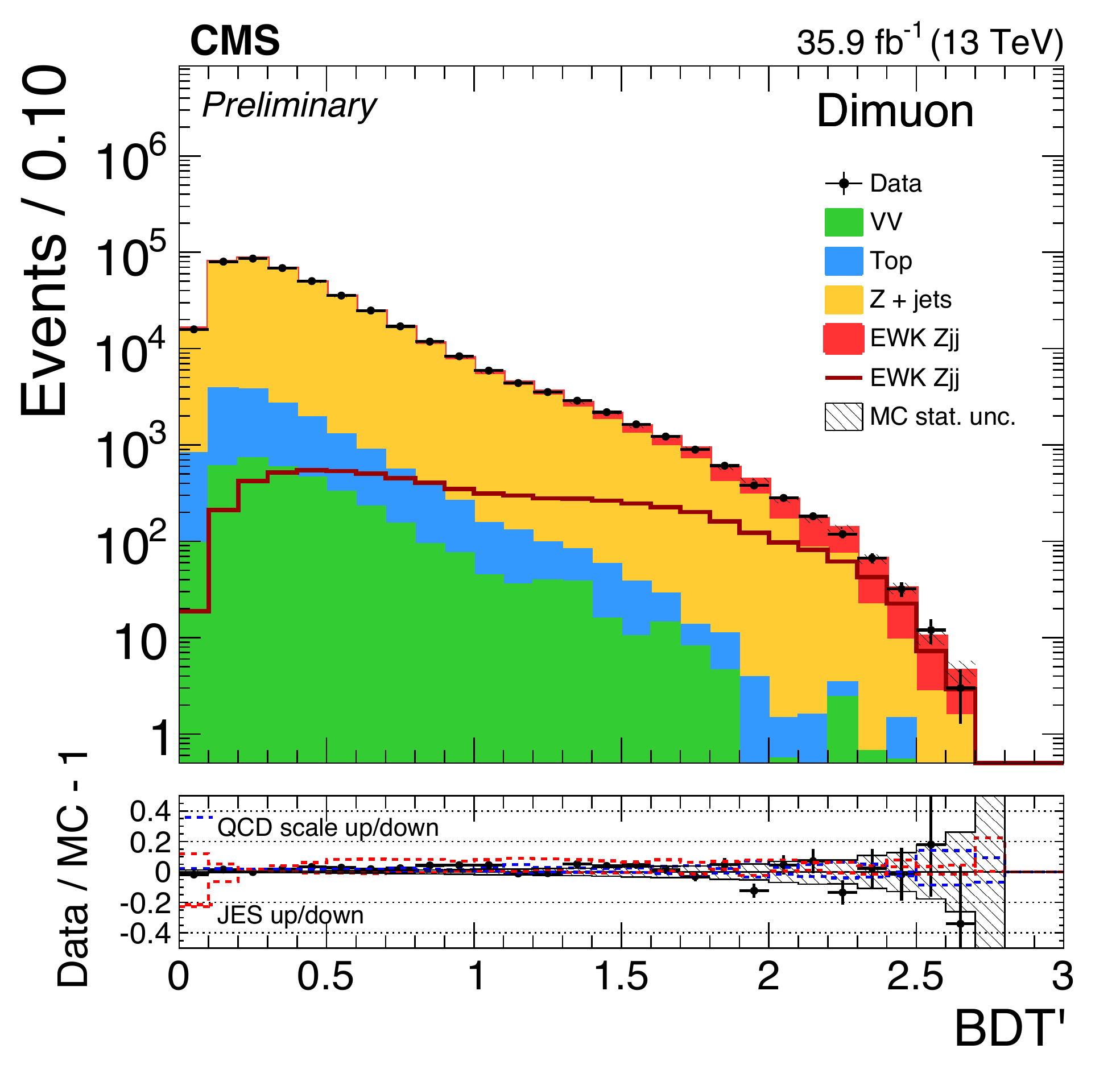}
\caption{(Left) Invariant mass of the dijet system for data and simulated events. 
The contributions from the different background sources and the signal are shown stacked, with data points superimposed. The expected signal-only contribution is also shown as an unfilled histogram. The lower panel shows the relative difference between the data and expectations as well as the uncertainty envelopes for JES and QCD scales uncertainties. 
(Right) Distributions for the BDT discriminants in dimuon events~\cite{PAS}. }
\label{fig:figure9}
\end{figure}

Predicted and observed distributions of the dijet invariant mass for events in the signal region is shown 
in Fig.~\ref{fig:figure8} left.
The measurement of the fiducial EWK Wjj cross section in the signal region is performed with an extended
joint binned likelihood fit of the dijet mass distribution for the normalization factors of the QCD and EWK Powheg+Pythia8 predictions.
The region at relatively low invariant mass 500--1000 GeV has low signal purity and primarily determines QCD contribution, while events
 with higher invariant mass have higher signal purity and mainly determine EWK contribution. 
The interference between the processes is not included in the fit, and is instead taken as an uncertainty based on SM predictions.
The measured fiducial EWK cross sections $144 \pm 23 (stat) \pm 23 (exp) \pm 13 (theo)$ fb for 7 TeV
 and $159 \pm 10 (stat) \pm 17 (exp) \pm 20 (theo)$ fb for 8 TeV can be compared to predicted values of $144 \pm 11$ and $198 \pm 12$ fb
for 7 and 8 TeV respectively. The paper also includes number of differential cross section measurements and 
set limits on anomalous triple-gauge-boson couplings. Figure~\ref{fig:figure8} right demonstrates  ratio of the measured to predicted values for
different measurements of the cross section times branching fractions of electroweak production of a
single W, Z, or Higgs boson at high dijet invariant mass for 7 ans 8 TeV. Within uncertainties the measurements
agree with predictions.

The above comparison does not yet include the first preliminary measurement of the EWK production of Z bosons at 13 TeV that was performed by CMS experiment~\cite{PAS}.
The measurement requires pairs of electrons or muons, with mass within 15 GeV from the nominal
 Z-boson mass for leptons with $p_T > 30(20)$ GeV, leading(subleading) and two jets with $p_T > 50$ and 30 GeV.
The dijet invariant mass for the dimuon channel is shown in Fig.~\ref{fig:figure9} left. One can see that the EWK relative 
contribution increases
with dijet mass as expected but still an order of magnitude less than the background QCD process even for very high masses. To improve the 
measurement a boosted decision tree (BDT) is used based on set of variables like the dijet pseudorapidity opening $\Delta \eta_{jj}$,
the dijet transverse momentum and others. The BDT is trained to achieve the best separation between EWK and DY production and the results 
for dimuon channel  are shown in 
Fig.~\ref{fig:figure9} right. By comparing the left to the right plot, one can obviously see that, as expected, the BDT provides better separation 
between the signal and the background as the dijet mass distribution alone.
The measured EWK cross section by combining electron and muon channels: $552 \pm 19 (stat) \pm 55 (syst)$ fb agrees well with 
the SM prediction, $543 \pm 24$ fb.

\section{Summary}

These proceedings present only some selected EWK measurements, there were more contributions presented 
 at this conference that covered other EWK topics. The precise
measurements require time to perform the data analyses  and therefore we only now complete 7+8 TeV program and start to analyse 13 TeV data. 
In most of the 
cases the statistical uncertainties do not dominate the precision of the measurements and better understanding of systematics, 
and in some cases theoretical uncertainties, are required. 

\vspace*{1cm}
The University of Wisconsin – Madison would like to acknowledge the U.S. Department of Energy (DOE) and the National Science Foundation (NSF) for funding their contribution to this research.

\end{document}